\documentclass[aps,prd,onecolumn,groupedaddress,showpacs,nofootinbib,amssymb]{revtex4}
\usepackage[dvips]{graphicx}
\usepackage{amssymb}
\usepackage{amsmath}
\usepackage{graphicx,,color}
\usepackage{amsfonts}
\usepackage{bm}
\usepackage{cancel}
\usepackage{comment}

\newcommand\be{\begin{equation}}
\newcommand\ee{\end{equation}}
\newcommand\nn{\nonumber \\}
\newcommand\e{\mathrm{e}}

\allowdisplaybreaks[4]

\begin{document}

\title{Ghost-Free $F(R)$ Gravity with Lagrange Multiplier Constraint}
\author{S.~Nojiri,$^{1,2,3}$\,\thanks{nojiri@gravity.phys.nagoya-u.ac.jp}
S.~D.~Odintsov,$^{4,5}$\,\thanks{odintsov@ieec.uab.es}
V.~K.~Oikonomou,$^{6,7}$\,\thanks{v.k.oikonomou1979@gmail.com}}
\affiliation{$^{1)}$ Department of Physics, Nagoya University,
Nagoya 464-8602, Japan \\
$^{2)}$ Kobayashi-Maskawa Institute for the Origin of Particles and
the Universe, Nagoya University, Nagoya 464-8602, Japan \\
$^{3)}$ KEK Theory Center, High Energy Accelerator Research Organization (KEK),
Oho 1-1, Tsukuba, Ibaraki 305-0801, Japan. \\
$^{4)}$ ICREA, Passeig Luis Companys, 23, 08010 Barcelona, Spain\\
$^{5)}$ Institute of Space Sciences (IEEC-CSIC) C. Can Magrans s/n,
08193 Barcelona, Spain\\
$^{6)}$ Laboratory for Theoretical Cosmology, Tomsk State University
of Control Systems
and Radioelectronics, 634050 Tomsk, Russia (TUSUR)\\
$^{7)}$ Tomsk State Pedagogical University, 634061 Tomsk, Russia\\
}

\tolerance=5000

\begin{abstract}
We propose two new versions of ghost-free
generalized $F(R)$ gravity with Lagrange multiplier constraint. The
first version of such theory for a particular degenerate choice of the
Lagrange multiplier, corresponds to mimetic $F(R)$ gravity. The
second version of such theory is  just the Jordan frame description
of mimetic gravity with potential. As we demonstrate, it is possible
to realize several cosmological scenarios in such theory. In particulary,
de Sitter solutions may also be found.
\end{abstract}


\maketitle

\section{Introduction}

Two of the most difficult to explain mysteries, that still haunt
modern theoretical cosmology, are the dark energy and dark matter
problems. The dark energy problem is related to the late-time
acceleration that the Universe experiences at present time, which
was firstly observed in the late 90's, while the dark matter problem
is a bit older, and various proposals have appeared in the
theoretical physics literature, that may model dark matter in an
adequate way. The most well-known description of dark matter is
given by particle physics, however modified gravity (for review, see
\cite{Nojiri:2006ri,Capozziello:2011et,Capozziello:2010zz,Capozziello:2009nq,Nojiri:2010wj,
Nojiri:2017ncd}) has appealing properties and even dark matter can
occur in a geometric way. A recent description of dark matter by
using a geometric approach, was offered by mimetic gravity
\cite{Sebastiani:2016ras}, which firstly appeared in
\cite{Chamseddine:2013kea} and later was developed for cosmological
purposes in \cite{Chamseddine:2014vna,Golovnev:2013jxa,
Deruelle:2014zza,Nojiri:2014zqa,Odintsov:2015wwp,Odintsov:2015ocy,
Odintsov:2015cwa,Leon:2014yua,Momeni:2015gka,
Astashenok:2015haa,Myrzakulov:2015qaa,Cognola:2016gjy,Ijjas:2016pad,Matsumoto:2015wja,
Bouhmadi-Lopez:2017lbx,
Mirzagholi:2014ifa,Sadeghnezhad:2017hmr,Gorji:2017cai,Zheng:2017qfs},
and also cosmological and astrophysical applications can be found in
\cite{Myrzakulov:2015kda,Oikonomou:2016fxb,Oikonomou:2016pkp,
Nojiri:2016ppu,Odintsov:2016imq}. In Ref.~\cite{Nojiri:2014zqa}, the
Einstein-Hilbert mimetic theoretical framework was extended in the
$F(R)$ mimetic gravity with Lagrange multiplier and potential. The
mimetic $F(R)$ gravity theory has many appealing properties and it
is possible to describe in a unified way inflation and the late-time
acceleration era \cite{Odintsov:2015wwp,Odintsov:2015ocy,
Odintsov:2015cwa}.

The purpose of this paper is to investigate certain theoretical
problems of generalized Lagrange-multiplier $F(R)$ gravity, with the
mimetic $F(R)$ gravity belonging to some subcases of the generalized
Lagrange-multiplier $F(R)$ gravity. Specifically, we shall be
interested in the appearance of ghosts in this kind of theories, and
we discuss the theoretical techniques that eliminate the ghosts. It is
known the mimetic $F(R)$ gravity theories with Lagrange
multiplier, always contain ghosts, thus making the theory less
appealing. However, as we demonstrate, the ghost may be eliminated
by appropriately modifying the gravitational action. Particularly,
we shall present two wide classes of generalized $F(R)$ gravity,
in which the ghosts are absent. In the context of these
cosmologies it is possible to realize various cosmological
scenarios, and we present some concrete examples.

This paper is organized as follows: In section II, we demonstrate in
detail how the ghost fields occur in the mimetic $F(R)$ gravity with
Lagrange multiplier, which is a degenerated subcase of the generalized
$F(R)$ gravity with Lagrange multipliers. In section III we present a
general class of $F(R)$ gravities with Lagrange multiplier, which
are free of ghosts. In section III, another ghost-free class of
models is introduced, and several cosmological realizations are
presented. Finally, the conclusions follow in the end of the paper.

Before we proceed, let us briefly mention the geometrical background
which shall be assumed in this paper, which is a flat
Friedmann-Robertson-Walker (FRW) geometric background, with line
element,
\begin{equation}
\label{JGRG14}
ds^2 = - dt^2 + a(t)^2 \sum_{i=1,2,3}
\left(dx^i\right)^2\, ,
\end{equation}
with $a(t)$ being the scale factor of the Universe. Also, the
connection is assumed to be a metric compatible, torsion-less and
symmetric connection, the Levi-Civita connection.

\section{Existence of Ghost Fields in Mimetic Lagrange Multiplier $F(R)$ Gravity: A Short Review}

In this section, we shall demonstrate that ghost fields exist in the
Lagrange multiplier $F(R)$ gravity, following
Ref.~\cite{Hirano:2017zox}. The action of the mimetic $F(R)$ gravity with
Lagrange multiplier is,
\begin{equation}
\label{FRmim1}
S_{F(R)}= \frac{1}{2\kappa^2}\int d^4 x \sqrt{-g} \left\{  F(R)
+ \lambda \left( \partial_\mu \phi \partial^\mu \phi + 1 \right) \right\}
\, ,
\end{equation}
where $\lambda$ is the Lagrange multiplier and $\phi$ is a scalar
field. Let us now demonstrate how ghost fields may emerge in such a
theory. The action of the mimetic $F(R)$ model (\ref{FRmim1}), can
be rewritten by introducing the auxiliary field $A$, as follows,
\begin{equation}
\label{JGRG21min}
S=\frac{1}{2\kappa^2}\int d^4 x \sqrt{-g} \left\{ F'(A)\left(R-A\right)
+ F(A)
+ \lambda \left( \partial_\mu \phi \partial^\mu \phi + 1 \right) \right\}\, .
\end{equation}
Upon variation of the action (\ref{JGRG21min}) with respect to $A$,
we obtain the relation $A=R$. Substituting $A=R$ into the action
(\ref{JGRG21min}), we can reproduce the action in
Eq.~(\ref{FRmim1}). Furthermore, upon conformally transforming the
metric in the following way,
\begin{equation}
\label{JGRG22}
g_{\mu\nu}\to \e^\sigma g_{\mu\nu}\, ,\quad \sigma = -\ln F'(A)\, ,
\end{equation}
the Einstein frame action is obtained,
\begin{align}
\label{JGRG23min}
S_E =& \frac{1}{2\kappa^2}\int d^4 x \sqrt{-g} \left\{ R
 - \frac{3}{2}g^{\rho\sigma}
\partial_\rho \sigma \partial_\sigma \sigma - V(\sigma)
+ \lambda \left( \e^\sigma \partial_\mu \phi \partial^\mu \phi
+ \e^{2\sigma} \right) \right\}\, ,\nn
V(\sigma) =& \e^\sigma g\left(\e^{-\sigma}\right)
 - \e^{2\sigma} f\left(g\left(\e^{-\sigma}\right)\right) =
\frac{A}{F'(A)}
 - \frac{F(A)}{F'(A)^2}\, .
\end{align}
In the above action, $g\left(\e^{-\sigma}\right)$ is given by
solving the equation $\sigma = -\ln\left( 1 + f'(A)\right)=- \ln
F'(A)$ as $A=g\left(\e^{-\sigma}\right)$. Upon variation of the
action with respect to $\lambda$, we obtain
\begin{equation}
\label{FRmim2}
\sigma = \ln \left( - \partial_\mu \phi
\partial^\mu \phi \right)\, .
\end{equation}
By substituting Eq.~(\ref{FRmim2})
in the action of Eq.~(\ref{JGRG23min}), we obtain,
\begin{equation}
\label{FRmim3}
S_E = \frac{1}{2\kappa^2}\int d^4 x \sqrt{-g}
\left\{ R - \frac{3}{2}
\left( - \partial_\mu \phi \partial^\mu \phi \right)^{-2}
g^{\rho\sigma}
\partial_\rho \left( - \partial_\mu \phi \partial^\mu \phi \right)
\partial_\sigma \left( - \partial_\mu \phi \partial^\mu \phi \right)
 - V(\sigma) \right\}\, .
\end{equation}
Due to the fact that higher derivative terms occur, the existence of
ghost fields is unavoidable.

In order to see this in a more rigid and formal way, we shall
investigate the perturbations around the flat Minkowski spacetime.
The existence of a Minkowski solution in Eq.~(\ref{FRmim1}) requires
that the following condition holds true,
\begin{equation}
\label{Minmim1}
V(\sigma) = V'(\sigma)=0 \, ,
\end{equation}
which implies,
\begin{equation}
\label{Minmim2}
F(A)=0 \, ,
\end{equation}
and in effect, the solution is given by
\begin{equation}
\label{Minmim3}
\lambda=0\, , \quad \phi = t \, .
\end{equation}
We consider the following perturbation,
\begin{equation}
\label{Minmim4}
\phi = t + \varphi \, ,
\end{equation}
and we expand the action with respect to $\varphi$. Then we find the
following Lagrangian,
\begin{equation}
\label{Minmim5}
\mathcal{L}_\varphi\equiv \frac{3}{2}
\left( - \partial_\mu \phi \partial^\mu \phi \right)^{-2}
g^{\rho\sigma}
\partial_\rho \left( - \partial_\mu \phi \partial^\mu \phi \right)
\partial_\sigma \left( - \partial_\mu \phi \partial^\mu \phi \right)
  - V(\sigma)
\sim - 6 \partial_\mu \left( \partial_t \varphi \right)
\partial^\mu \left( \partial_t \varphi \right)
 - \alpha \left( \partial_t \varphi \right)^2 \, .
\end{equation}
where $\alpha$ stands for,
\begin{equation}
\label{Minmim6}
\alpha \equiv \frac{F'''(0)}{2 F''(0)^2} - \frac{6}{F''(0)} \, .
\end{equation}
Since we are interested in the existence of ghost fields, we neglect
the spatial derivatives in the above Lagrangian, and by introducing
a new scalar field $\eta$, we rewrite the Lagrangian
$\mathcal{L}_{\varphi}$ as follows,
\begin{equation}
\label{Minmim7}
\mathcal{L}_\varphi \sim - 6 \partial_t \eta \partial_t \varphi
 - \alpha \left( \partial_t \varphi \right)^2 - 6 \lambda^2 \, .
\end{equation}
The matrix $M$ which is composed by the coefficients of the kinetic
terms in the Lagrangian (\ref{Minmim7}), are given by,
\begin{equation}
\label{Minmim8}
M = \left( \begin{array}{cc} - \alpha & -3 \\ -3 & 0 \end{array} \right)
\, .
\end{equation}
Due to the fact that the determinant of the matrix $M$ is negative
$\det M = - 9$, one of the two eigenvalues is always positive, but
the other is always negative, and  therefore the ghost fields
definitely exist in the theory.

\section{Ghost-free Generalized Lagrange Multiplier $F(R)$ gravity: Model I}

As we explicitly demonstrated in the previous section, ghost fields
appear in the model of Eq.~(\ref{FRmim1}), so  in this section we
shall consider a variant form of the model (\ref{FRmim1}), in which
no ghosts appear. The modified model has the following action,
\begin{equation}
\label{FRmim4}
S_{F(R)}= \frac{1}{2\kappa^2}\int d^4 x \sqrt{-g} \left\{  F(R)
+ \lambda \left( \partial_\mu \phi \partial^\mu \phi + G(R) \right) \right\}
\, ,
\end{equation}
where $G(R)$ is an differentiable function of the scalar curvature
$R$. As we did in the case of the action (\ref{JGRG21min}), we
rewrite the action (\ref{FRmim4}) as follows,
\begin{equation}
\label{FRmim5}
S=\frac{1}{2\kappa^2}\int d^4 x \sqrt{-g} \left\{
\left( F'(A) + \lambda G'(A) \right)\left(R-A\right)
+ F(A)
+ \lambda \left( \partial_\mu \phi \partial^\mu \phi + G(A) \right) \right\}\, .
\end{equation}
Again upon varying the action with respect to $A$, we obtain the
equation $A=R$. Substituting $A=R$ in the action of
Eq.~(\ref{FRmim5}), we reproduce the action of Eq.~(\ref{FRmim4}).
Instead of (\ref{JGRG22}), by using the following conformal
transformation,
\begin{equation}
\label{JGRG22mim}
g_{\mu\nu}\to \e^\sigma g_{\mu\nu}\, ,\quad
\sigma = - \ln \left( F'(A) + \lambda G'(A) \right) \, ,
\end{equation}
we obtain the following Einstein frame action,
\begin{align}
\label{FRmim6}
S_E =& \frac{1}{2\kappa^2}\int d^4 x \sqrt{-g}
\left\{ R - \frac{3}{2}g^{\rho\sigma}
\partial_\rho \sigma \partial_\sigma \sigma - V(\sigma)
+ \lambda \left( \e^\sigma \partial_\mu \phi \partial^\mu \phi
+ \e^{2\sigma} G(A) \right) \right\}\, ,\nn
V(\sigma) =& \frac{A}{F'(A)} - \frac{F(A)}{F'(A)^2}\, .
\end{align}
By using the second equation in (\ref{JGRG22mim}), we may eliminate
the function $\lambda$ as long as the condition $G'(A)\neq 0$ holds
true, and  we obtain,
\begin{align}
\label{FRmim6B}
S_E =& \frac{1}{2\kappa^2}\int d^4 x \sqrt{-g}
\left\{ R - \frac{3}{2}g^{\rho\sigma}
\partial_\rho \sigma \partial_\sigma \sigma - V(A,\sigma)
+ \frac{\e^{-\sigma} - F'(A)}{G'(A)}
\left( \e^\sigma \partial_\mu \phi \partial^\mu \phi
+ \e^{2\sigma} G(A) \right) \right\}\, ,\nn
V(A,\sigma) =& A \e^\sigma - F(A) \e^{2\sigma}\, .
\end{align}
We should note that the model (\ref{FRmim1}) corresponds to $G(A)=1$
and therefore $G'(A)=0$. By using the equation obtained when the
action is varied with respect to  $A$,
\begin{equation}
\label{FRmin6C}
0= \left( - \frac{F''(A)}{G'(A)} - \frac{\left(\e^{-\sigma} - F'(A)\right)G''(A)}{G'(A)^2}
\right) \left( \e^\sigma \partial_\mu \phi \partial^\mu \phi
+ \e^{2\sigma} G(A) \right) \, ,
\end{equation}
we can find the function $A$
as a function of $\sigma$ and $\partial_\mu \phi \partial^\mu \phi$
as $A=A\left(\sigma, \partial_\mu \phi
\partial^\mu \phi\right)$. In effect, Eq.~(\ref{FRmim6B}) can be written as follows,
\begin{align}
\label{FRmim7}
S_E =& \frac{1}{2\kappa^2}\int d^4 x \sqrt{-g}
\left[ R - \frac{3}{2}g^{\rho\sigma} \partial_\rho \sigma
\partial_\sigma \sigma - V\left( A\left(\sigma, \partial_\mu \phi \partial^\mu \phi\right),
 \sigma\right)  \right. \nn
& \left. + \frac{\e^{-\sigma} - F'\left(
A\left(\sigma, \partial_\mu \phi \partial^\mu \phi\right) \right)}
{G'\left( A\left(\sigma, \partial_\mu \phi \partial^\mu \phi\right)
\right)} \left( \e^\sigma \partial_\mu \phi \partial^\mu \phi
+ \e^{2\sigma} G\left( A\left(\sigma, \partial_\mu \phi \partial^\mu
\phi\right) \right) \right) \right]\, .
\end{align}
Although it is difficult to find the explicit form of
$A\left(\sigma, \partial_\mu \phi \partial^\mu \phi\right)$, the
action (\ref{FRmim7}) does not include any higher derivative terms,
a result which is different from the one corresponding in the action
of Eq.~(\ref{FRmim3}), as we showed in the previous section.
Therefore the model of Eq.~(\ref{FRmim4}) is ghost free.

If $ - \frac{F''(A)}{G'(A)} - \frac{\left(\e^{-\sigma} - F'(A)\right)G''(A)}{G'(A)^2}
\neq 0$, Eq.~(\ref{FRmin6C}) gives,
\begin{equation}
\label{FRmin7b}
0 = \e^\sigma \partial_\mu \phi \partial^\mu \phi
+ \e^{2\sigma} G(A) \, ,
\end{equation}
which is nothing but the constraint equation in the Einstein frame
given by the variation of $\lambda$ in the Jordan frame action (\ref{FRmim5}).
The variations of the action (\ref{FRmim6B}) with respect to $\sigma$, $\phi$, and $g_{\mu\nu}$
give
\begin{align}
\label{FRmin7c}
0 = & \frac{3}{2} \nabla_\mu \nabla^\mu \sigma -  A \e^\sigma + 2 F(A) \e^{2\sigma}
 -  \frac{\e^{-\sigma}}{G'(A)}
\left( \e^\sigma \partial_\mu \phi \partial^\mu \phi
+ \e^{2\sigma} G(A) \right)
+ \frac{\e^{-\sigma} - F'(A)}{G'(A)}
\left( \e^\sigma \partial_\mu \phi \partial^\mu \phi
+ 2 \e^{2\sigma} G(A) \right) \, , \\
\label{FRmin7d}
0 = &  \nabla^\mu \left( \frac{\e^{-\sigma} - F'(A)}{G'(A)}
\e^\sigma \partial_\mu \phi \right) \, , \\
\label{FRmin7e}
0 = & - R_{\mu\nu} + \frac{1}{2} g_{\mu\nu} R
+ \frac{1}{2} \left\{ - \frac{3}{2}g^{\rho\sigma}
\partial_\rho \sigma \partial_\sigma \sigma - V(A,\sigma)
+ \frac{\e^{-\sigma} - F'(A)}{G'(A)}
\left( \e^\sigma \partial_\mu \phi \partial^\mu \phi
+ \e^{2\sigma} G(A) \right) \right\} g_{\mu\nu} \nn
& + \frac{3}{2} \partial_\mu \sigma \partial_\nu \sigma
 - \frac{\left( \e^{-\sigma} - F'(A) \right) \e^\sigma}{G'(A)}
\partial_\mu \phi \partial_\nu \phi \, , \\
\end{align}
We now consider the condition that the flat Minkowski space-time becomes a solution.
Because $A$ is nothing but the scalar curvature in the Jordan frame, we require $A=0$ and
we also assume that $\sigma$ is a constant and $\phi$ only depends on time $t$,
\begin{equation}
\label{FRmin7a1}
A=0\, , \quad \sigma = \sigma_0 \, , \quad \phi=\phi(t) \, .
\end{equation}
Then Eq.~(\ref{FRmin7d}) is trivially satisfied and Eqs.~(\ref{FRmin7b}), (\ref{FRmin7c}), and
(\ref{FRmin7e}) reduce to the following forms,
\begin{align}
\label{FRmin7a2}
0 =& - \e^{\sigma_0} {\dot\phi}^2 + \e^{2\sigma_0} G(0) \, , \\
\label{FRmin7a3}
0 = & 2 F(0) \e^{2\sigma_0} + \frac{\e^{-\sigma_0} - F'(0)}{G'(0)}
\left( - \e^{\sigma_0} {\dot\phi}^2 + 2 \e^{2\sigma_0} G(0) \right) \, , \\
\label{FRmin7a4}
0 = & - \frac{1}{2} F(0) \e^{2\sigma_0} -  \frac{\e^{-\sigma_0} - F'(0)}{G'(0)}
\e^{\sigma_0} {\dot\phi}^2 \\
\label{FRmin7a4B}
0 = &  \frac{1}{2} F(0) \e^{2\sigma_0} \, .
\end{align}
Then by using (\ref{FRmin7a2}), we find
\begin{equation}
\label{FRmin7a7}
0 = F(0) = \left( \e^{-\sigma_0} - F'(0)\right) G(0) \, ,
\end{equation}
Eq.~(\ref{FRmin7a2}) can be solved to give
\begin{equation}
\label{FRmin7a8}
\phi = \phi_0 \pm t \e^{\frac{\sigma_0}{2}} \sqrt{G_0} \, .
\end{equation}
Here $\phi_0$ is a constant.
In order to investigate if there is a ghost or not, we consider the perturbation from the flat
Minkowski space-time.
By using (\ref{FRmin7a7}), we consider the case that $F(0)=G(0)=0$ and the perturbation
from the solution given by (\ref{FRmin7a1}) and (\ref{FRmin7a8}),
\begin{equation}
\label{FRmin7a9}
A=\delta A \, , \quad \sigma = \sigma_0 + \delta \sigma \, , \quad
\phi = \phi_0 + \delta \phi \, .
\end{equation}
Then the scalar part in the action (\ref{FRmim6B}) has the following form,
\begin{align}
\label{FRmim6B1}
S_E =& \frac{1}{2\kappa^2}\int d^4 x \sqrt{-g}
\left\{ R - \frac{3}{2} \partial_\mu \delta \sigma \partial^\mu \delta \sigma
 - \e^{\sigma_0} \delta A \delta\sigma + 2 \e^{2\sigma_0} F'(0) \delta A \delta \sigma
\right. \nn
& + \frac{\left( \e^{-\sigma_0} - F'(0) \right)\e^{\sigma_0}}{G'(0)}
\left( \partial_\mu \delta \phi \partial^\mu \delta \phi
+ \frac{1}{2} \e^{2\sigma_0} G''(0) \delta A^2
+ 2 \e^{2\sigma_0} G'(0) \delta \sigma \delta A \right) \nn
& \left. + \left( - \e^{ - \sigma_0} \delta \sigma - F''(0) \delta A
 - \frac{\left( \e^{-\sigma_0} - F'(0) \right) G''(0)}{G'(0)} \delta A
\right) \e^{2\sigma_0} \delta A \right\}\, .
\end{align}
The equation given by the variation with respect to $\delta A$ gives
$\delta A$ in terms of $\sigma$. Then by substituting the expression
$\delta A = C \delta \sigma$ with a constant $C$, we obtain the mass
term for $\delta \sigma$. The action (\ref{FRmim6B1}) tells that as
long as the following relation holds true,
\begin{equation}
\label{FRmim6B2}
\frac{\e^{-\sigma_0} - F'(0)}{G'(0)}<0 \, ,
\end{equation}
the ghost does not appear.

By varying the action (\ref{FRmim4}) with respect to the function
$\lambda$ and with respect to the scalar field $\phi$, we obtain the
following equations,
\begin{align}
\label{mimmim1}
0 =& \partial_\mu \phi \partial^\mu \phi + G(R) \, , \\
\label{mimmim2}
0 =& \nabla^\mu \left( \lambda \partial_\mu  \phi \right)\, ,
\end{align}
On the other hand, upon variation of the action with respect to the
metric $g_{\mu\nu}$, we obtain,
\begin{equation}
\label{mimmim3}
0 = \frac{F(R)}{2} g_{\mu\nu} - \left( F'(R) + \lambda G'(R) \right)
R_{\mu\nu} - \lambda \partial_\mu \phi \partial_\nu \phi
+ \left(\nabla_\mu \nabla_\nu - g_{\mu\nu} \nabla^2 \right)
\left( F'(R) + \lambda G'(R) \right) \, .
\end{equation}

Let us now demonstrate how the gravitational equations of the model
(\ref{FRmim4}) become, when a specific cosmological background is
considered. We assume that the geometric background is flat a FRW
metric with line element of the form of Eq.~(\ref{JGRG14}), and also
that the function $\lambda$ and also the scalar field $\phi$ depend
only on the cosmic time $t$. In effect, the Eqs.~(\ref{mimmim1}) and
(\ref{mimmim2}), take the following form,
\begin{equation}
\label{mimmim4}
0 = - {\dot\phi}^2 + G(R) \, , \quad
0 = \frac{d}{dt} \left( a^3 \lambda \dot \phi \right) \, ,
\end{equation}
which can be rewritten as follows,
\begin{equation}
\label{mimmim5}
\dot\phi = \pm \sqrt{G(R)} \, , \quad
a^3 \lambda \dot \phi = C \, ,
\end{equation}
where $C$ is an integration constant. Also, the $(t,t)$ and $(i,j)$
components of Eq.~(\ref{mimmim3}) yield the following equations,
\begin{align}
\label{mimmim6}
0 = & - \frac{F(R)}{2}
+ 3 \left( \dot H + H^2 \right) \left( F'(R) \pm \frac{C G'(R)}{a^3 \sqrt{G(R)}} \right)
\mp \frac{C \sqrt{G(R)}}{a^3} - 3 H \frac{d}{dt}
\left( F'(R) \pm \frac{C G'(R)}{a^3 \sqrt{G(R)}} \right) \, , \\
\label{mimmim7}
0 = & \frac{F(R)}{2} - \left( \dot H + 3 H^2 \right)
\left( F'(R) \pm \frac{C G'(R)}{a^3 \sqrt{G(R)}} \right)
+ \left( \frac{d^2}{dt^2} + 2 H \frac{d}{dt} \right)
\left( F'(R) \pm \frac{C G'(R)}{a^3 \sqrt{G(R)}} \right) \, .
\end{align}
If we define a new quantity $J(R,a)$ as follows,
\begin{equation}
\label{mimmim8}
J(R,a) \equiv F(R) \pm \frac{2C \sqrt{G(R)}}{a^3} \, ,
\end{equation}
Eq.~(\ref{mimmim6}) can be rewritten in the following form,
\begin{equation}
\label{mimmim9}
0 = - \frac{J(R,A)}{2}
+ 3 \left( \dot H + H^2  - 3 H \frac{d}{dt} \right) \frac{\partial J(R,a)}{\partial R} \, .
\end{equation}
We should note that when $C=0$, Eqs.~(\ref{mimmim6}) and
(\ref{mimmim7}) become identical to the equations of the standard
$F(R)$ gravity, which indicates that any solution of the standard
$F(R)$ gravity is also a solution of the model (\ref{FRmim4}).

An analytic form for the $F(R)$ and $G(R)$ gravity, can be given if
the de Sitter spacetime is considered, in which case $H=H_0$ and
$a=\e^{H_0 t}$. Then Eqs.~(\ref{mimmim6}) and (\ref{mimmim7}) can be
cast in the following form,
\begin{align}
\label{mimmim10}
0 = & - \frac{F\left( R_0 \right)}{2}
+ 3 H_0^2 F'(R_0)
\pm \frac{C}{a^3 \sqrt{G(R_0)}}  \left( 12 H_0^2 G'(R_0) - G(R_0) \right) \, , \\
\label{mimmim11} 0 = & \frac{F(R_0)}{2} - 3 H_0^2 F'(R_0) \, ,
\end{align}
where $R_0=12 H_0^2$. Then in order for the solution describing the
de Sitter space-time to exist, the functions $F(R)$ and $G(R)$ must
simultaneously satisfy the following differential equations,
\begin{equation}
\label{mimmim12}
0 = 2 F(R_0) - R_0 F'(R_0) \, , \quad 0 = R_0 G'(R_0) - G(R_0) \, .
\end{equation}
A special solution to the differential equations (\ref{mimmim12}) is
the following,
\begin{equation}
\label{mimmim13} F(R) =\alpha R^2 \, , \quad G(R) =\beta R \, ,
\end{equation}
and both the differential equations(\ref{mimmim12}) are satisfied.
Note that other examples of such theory leading to de Sitter space maybe
found.

\section{Ghost-free Generalized  $F(R)$ gravity: Model II}

Another ghost-free model of generalized $F(R)$ gravity, can be
obtained in the Einstein frame, if the scalar fields $\tilde\lambda$
and $\phi$ are introduced in the Lagrangian as follows
\cite{Chamseddine:2014vna},
\begin{equation}
\label{FRmin2}
S_E = \frac{1}{2\kappa^2}\int d^4 x \sqrt{-g}
\left\{ R - \frac{3}{2}g^{\rho\sigma}
\partial_\rho \sigma \partial_\sigma \sigma - V(\sigma)
+ \tilde \lambda \left( \partial_\mu \phi \partial^\mu \phi + 1 \right)
\right\} \, .
\end{equation}
By applying the inverse of the transformation (\ref{JGRG22}), we
obtain,
\begin{equation}
\label{FRmin3} S=\frac{1}{2\kappa^2}\int d^4 x \sqrt{-g}
\left\{F'(A)\left(R-A\right) + F(A) + \lambda \left( \partial_\mu
\phi \partial^\mu \phi + F'(A) \right) \right\}\, ,
\end{equation}
where $\lambda = F'(A) \tilde \lambda$. Upon varying the action with
respect to $A$, we obtain the following equation,
\begin{equation}
\label{FRmin4}
A = R + \lambda \, .
\end{equation}
Then by substituting Eq.~(\ref{FRmin4}) in the action
(\ref{FRmin3}), we obtain the following action,
\begin{equation}
\label{FRmim5C}
S_{F(R)}= \frac{1}{2\kappa^2}\int d^4 x \sqrt{-g} \left\{  F(R+\lambda)
+ \lambda \partial_\mu \phi \partial^\mu \phi \right\}
\, ,
\end{equation}
which is the action of the mimetic $F(R)$ gravity without ghost. If
we further redefine $\lambda$ as follows $\lambda \to \lambda - R$,
we obtain the following action,
\begin{equation}
\label{FRmim5B}
S_{F(R)}= \frac{1}{2\kappa^2}\int d^4 x \sqrt{-g} \left\{  F(\lambda)
+ \left(\lambda - R \right) \partial_\mu \phi \partial^\mu \phi \right\}
\, ,
\end{equation}
If we assume that the leading order of $F(\lambda)$ is linear,
\begin{equation}
\label{FRmimmim1}
F(\lambda) = \lambda + \mathcal{O}\left( \lambda^2 \right)\, ,
\end{equation}
or equivalently,
\begin{equation}
\label{FRmimmim2}
F(R+\lambda) = R+\lambda + \mathcal{O}\left( R+\lambda^2 \right)\, ,
\end{equation}
the leading order in the action (\ref{FRmim5C}) is effectively the
standard Einstein action with the mimetic constraint,
\begin{equation}
\label{FRmimmim3}
S = \frac{1}{2\kappa^2}\int d^4 x \sqrt{-g}
\left\{  R + \lambda \left( \partial_\mu \phi \partial^\mu \phi +1
\right) + \mathcal{O}\left( R+\lambda^2 \right) \right\} \, .
\end{equation}

The variation of the action (\ref{FRmim5B}) with respect to the
scalar fields $\lambda$, $\phi$ and also with respect to the metric
$g_{\mu\nu}$, gives the following equations,
\begin{align}
\label{FRmin6}
0 = & F'(\lambda) + \partial_\mu \phi \partial^\mu \phi \, , \\
\label{FRmin7}
0= & \nabla_\mu \left\{ \left(\lambda - R \right) \partial^\mu \phi \right\} \, ,
\\
\label{FRmin8}
0= & \frac{1}{2} \left\{  F(\lambda)
+ \left(\lambda - R \right) \partial_\mu \phi \partial^\mu \phi \right\}
g_{\mu\nu} + R_{\mu\nu} \partial_\rho \phi \partial^\rho
\phi - \nabla_\mu \nabla_\nu \left( \partial_\rho \phi \partial^\rho \phi \right)
+ g_{\mu\nu} \nabla^2 \left( \partial_\rho \phi \partial^\rho \phi \right) \, .
\end{align}
By assuming a spatially flat FRW universe (\ref{JGRG14}) and also
that $\lambda$ and $\phi$ depend  only on the cosmic time coordinate
$t$, the above equations (\ref{FRmin6}), (\ref{FRmin7}), and
(\ref{FRmin8}) take the following form,
\begin{align}
\label{FRmin9}
0 = & F'(\lambda) - {\dot \phi}^2 \, , \\
\label{FRmin10}
0= & \frac{d}{dt} \left\{ a^3 \left(\lambda - 6 \dot H - 12 H^2 \right)
\dot\phi \right\} \, ,
\\
\label{FRmin11}
0= & - \frac{1}{2} \left\{
F(\lambda) - \left(\lambda - 6 H^2 \right) {\dot\phi}^2
\right\} - 3 H \frac{d}{dt} \left( {\dot\phi}^2 \right) \, .
\\
\label{FRmin12}
0= & \frac{1}{2} \left\{
F(\lambda) - \left(\lambda - 4 \dot H - 6 H^2 \right) {\dot\phi}^2 \right\}
+  \left( \frac{d^2}{dt^2} + 4 H \frac{d}{dt} \right)
\left( {\dot\phi}^2 \right) \, .
\end{align}
In effect, Eq.~(\ref{FRmin10}) can be integrated and it yields,
\begin{equation}
\label{FRmin13}
C = a^3 \left(\lambda - 6 \dot H - 12 H^2 \right) \dot\phi \, ,
\end{equation}
with $C$ being again a constant of the integration. By using
Eq.~(\ref{FRmin9}), we may eliminate $\dot\phi$ from
Eqs.~(\ref{FRmin11}), (\ref{FRmin12}),  and (\ref{FRmin13}) as follows,
\begin{align}
\label{FRmin14}
0= & - \frac{1}{2} \left\{
F(\lambda) - \left(\lambda - 6 H^2 \right) F'(\lambda)
\right\} - 3 H \frac{d F'(\lambda)}{dt} \, . \\
\label{FRmin15}
0= & \frac{1}{2} \left\{
F(\lambda) - \left(\lambda - 4 \dot H - 6 H^2 \right) F'(\lambda) \right\}
+  \left( \frac{d^2 F'(\lambda)}{dt^2}
+ 4 H \frac{d F'(\lambda)}{dt} \right) \, , \\
\label{FRmin16}
C^2 =& a^6 \left(\lambda - 6 \dot H - 12 H^2 \right)^2 F'(\lambda) \, .
\end{align}
By using Eqs.~(\ref{FRmin14}) and (\ref{FRmin15}), we may eliminate
$F(\lambda)$ and we obtain,
\begin{equation}
\label{FRmin17}
0 = \frac{d^2 F'(\lambda)}{dt^2} + H \frac{d F'(\lambda)}{dt}
+ 2 \dot H F'(\lambda) \, .
\end{equation}
Let us now use the formalism we just presented in order to relaized
various cosmological scenarios. We shall assume that $C=0$ in
Eqs.~(\ref{FRmin13}) and (\ref{FRmin16}). Then we obtain,
\begin{equation}
\label{FRmin18}
\lambda = 6 \dot H + 12 H^2 \, .
\end{equation}
First we consider the power-law expansion of the Universe, in which
case the Hubble rate is,
\begin{equation}
\label{FRmin19}
a \propto t^{h_0}\quad \mbox{or} \quad H=\frac{h_0}{t}\, .
\end{equation}
Then Eq.~(\ref{FRmin18}) yields,
\begin{equation}
\label{FRmin20}
\lambda = \frac{ - 6 h_0 + 12h_0^2}{t^2} \, .
\end{equation}
On the other hand, by assuming $F'(\lambda) \propto t^{f_0}$,
Eq.~(\ref{FRmin17}) yields,
\begin{equation}
\label{FRmin21}
0 = f_0^2 + \left( h_0 - 1 \right) f_0 - 2h_0 \, ,
\end{equation}
which when solved yields,
\begin{equation}
\label{FRmin22}
f_0 = \frac{ 1 - h_0 \pm \sqrt{ h_0^2 +6 h_0 + 1}}{2} \, .
\end{equation}
Therefore $F'(\lambda) \propto \lambda^{- \frac{f_0}{2}}$ and
therefore we have,
\begin{equation}
\label{FRmin23}
F(\lambda) \propto \lambda^{1 - \frac{f_0}{2}}
= \lambda^{\frac{ 1 + h_0 \mp \sqrt{ h_0^2 +6 h_0 + 1}}{2} }\, .
\end{equation}
Now let us consider the de Sitter space-time, where $H$ is a
constant $H=H_0$, by assuming $C=0$, again. Then Eq.~(\ref{FRmin18})
gives,
\begin{equation}
\label{FRmin24}
\lambda = 12 H_0^2 \, .
\end{equation}
Since $\lambda$ is a constant, $F(\lambda)$ and $F'(\lambda)$ are
also constants. Then Eq.~(\ref{FRmin14}) or Eq.~(\ref{FRmin15})
yields,
\begin{equation}
\label{FRmin25}
0= F(\lambda) - \left( \lambda - 6 H_0^2 \right) F'(\lambda)
= F(\lambda) - \frac{\lambda}{2} F'(\lambda) \, ,
\end{equation}
which can be integrated and yields
\begin{equation}
\label{FRmin26}
F(\lambda) \propto \lambda^2 \, .
\end{equation}
Hence, we demonstrated that even in this ghost-free model, various
cosmological scenarios can be realized.

\section{Conclusions}

In this paper we presented two ghost-free generalized $F(R)$ gravity
models, which are variants to mimetic $F(R)$ gravity models, without
the ghost fields. After we discussed how ghosts may occur
in the mimetic $F(R)$ gravity models, we presented the first model,
which is a Lagrange multiplier $F(R)$ gravity model. Also we presented a
second model of standard mimetic gravity for which the Lagrangian
is considered in the Einstein and in the Jordan frame, and we demonstrated
that several cosmological scenarios can be realized.

In principle,  other cosmological solutions maybe constructed,
such as bouncing cosmologies or even alternative inflationary
scenarios, and also singular inflationary scenarios. We shall
address these issues in a future work.

\section*{Acknowledgments}

This work is supported by MINECO (Spain), project
FIS2016-76363-P and by CSIC I-LINK1019 Project (S.D.O) and (in part)
by MEXT KAKENHI Grant-in-Aid for Scientific Research on Innovative
Areas ``Cosmic Acceleration'' (No. 15H05890) (S.N.).


\begin{thebibliography}{99}

\bibitem{Nojiri:2006ri}
S.~Nojiri and S.~D.~Odintsov,
eConf C {\bf 0602061} (2006) 06 [Int.\ J.\ Geom.\ Meth.\ Mod.\
Phys.\ {\bf 4} (2007) 115] doi:10.1142/S0219887807001928
[hep-th/0601213].

\bibitem{Capozziello:2011et}
S.~Capozziello and M.~De Laurentis,
Phys.\ Rept.\ {\bf 509} (2011) 167 doi:10.1016/j.physrep.2011.09.003
[arXiv:1108.6266 [gr-qc]].

\bibitem{Capozziello:2010zz}
V.~Faraoni and S.~Capozziello,
Fundam.\ Theor.\ Phys.\ {\bf 170} (2010).
doi:10.1007/978-94-007-0165-6

\bibitem{Capozziello:2009nq}
S.~Capozziello, M.~De Laurentis and V.~Faraoni,
Open Astron.\ J.\ {\bf 3} (2010) 49 doi:10.2174/1874381101003010049,
10.2174/1874381101003020049 [arXiv:0909.4672 [gr-qc]].

\bibitem{Nojiri:2010wj}
S.~Nojiri and S.~D.~Odintsov,
Phys.\ Rept.\ {\bf 505} (2011) 59 doi:10.1016/j.physrep.2011.04.001
[arXiv:1011.0544 [gr-qc]].

\bibitem{Nojiri:2017ncd}
S.~Nojiri, S.~D.~Odintsov and V.~K.~Oikonomou,
Phys.\ Rept.\  {\bf 692} (2017) 1
doi:10.1016/j.physrep.2017.06.001
[arXiv:1705.11098 [gr-qc]].

\bibitem{Sebastiani:2016ras}
L.~Sebastiani, S.~Vagnozzi and R.~Myrzakulov,
arXiv:1612.08661 [gr-qc].

\bibitem{Chamseddine:2013kea}
A.~H.~Chamseddine and V.~Mukhanov,
JHEP {\bf 1311} (2013) 135
[arXiv:1308.5410 [astro-ph.CO]].

\bibitem{Chamseddine:2014vna}
A.~H.~Chamseddine, V.~Mukhanov and A.~Vikman,
JCAP {\bf 1406} (2014) 017
doi:10.1088/1475-7516/2014/06/017
[arXiv:1403.3961 [astro-ph.CO]].

\bibitem{Golovnev:2013jxa}
A.~Golovnev,
Phys.\ Lett.\ B {\bf 728} (2014) 39
[arXiv:1310.2790 [gr-qc]].

\bibitem{Deruelle:2014zza}
N.~Deruelle and J.~Rua,
JCAP {\bf 1409} (2014) 002
[arXiv:1407.0825 [gr-qc]].

\bibitem{Nojiri:2014zqa}
S.~Nojiri and S.~D.~Odintsov,
Mod.\ Phys.\ Lett.\ A {\bf 29} (2014) no.40, 1450211
doi:10.1142/S0217732314502113
[arXiv:1408.3561 [hep-th]].

\bibitem{Odintsov:2015wwp}
S.~D.~Odintsov and V.~K.~Oikonomou,
Phys.\ Rev.\ D {\bf 93} (2016) no.2, 023517
doi:10.1103/PhysRevD.93.023517
[arXiv:1511.04559 [gr-qc]].

\bibitem{Odintsov:2015ocy}
S.~D.~Odintsov and V.~K.~Oikonomou,
Astrophys.\ Space Sci.\ {\bf 361} (2016) no.5, 174
doi:10.1007/s10509-016-2761-9
[arXiv:1512.09275 [gr-qc]].

\bibitem{Odintsov:2015cwa}
S.~D.~Odintsov and V.~K.~Oikonomou,
Annals Phys.\ {\bf 363} (2015) 503
doi:10.1016/j.aop.2015.10.013
[arXiv:1508.07488 [gr-qc]].

\bibitem{Leon:2014yua}
G.~Leon and E.~N.~Saridakis,
JCAP {\bf 1504} (2015) no.04, 031
doi:10.1088/1475-7516/2015/04/031
[arXiv:1501.00488 [gr-qc]].

\bibitem{Momeni:2015gka}
D.~Momeni, R.~Myrzakulov and E.~Gudekli,
Int.\ J.\ Geom.\ Meth.\ Mod.\ Phys.\ {\bf 12} (2015) no.10, 1550101
doi:10.1142/S0219887815501017
[arXiv:1502.00977 [gr-qc]].

\bibitem{Astashenok:2015haa}
A.~V.~Astashenok, S.~D.~Odintsov and V.~K.~Oikonomou,
Class.\ Quant.\ Grav.\ {\bf 32} (2015) no.18, 185007
doi:10.1088/0264-9381/32/18/185007
[arXiv:1504.04861 [gr-qc]].

\bibitem{Myrzakulov:2015qaa}
R.~Myrzakulov, L.~Sebastiani and S.~Vagnozzi,
Eur.\ Phys.\ J.\ C {\bf 75} (2015) 444
doi:10.1140/epjc/s10052-015-3672-6
[arXiv:1504.07984 [gr-qc]].

\bibitem{Cognola:2016gjy}
G.~Cognola, R.~Myrzakulov, L.~Sebastiani, S.~Vagnozzi and S.~Zerbini,
Class.\ Quant.\ Grav.\ {\bf 33} (2016) no.22, 225014
doi:10.1088/0264-9381/33/22/225014
[arXiv:1601.00102 [gr-qc]].

\bibitem{Ijjas:2016pad}
A.~Ijjas, J.~Ripley and P.~J.~Steinhardt,
Phys.\ Lett.\ B {\bf 760} (2016) 132
doi:10.1016/j.physletb.2016.06.052
[arXiv:1604.08586 [gr-qc]].

\bibitem{Mirzagholi:2014ifa}
L.~Mirzagholi and A.~Vikman,
JCAP {\bf 1506} (2015) 028
doi:10.1088/1475-7516/2015/06/028
[arXiv:1412.7136 [gr-qc]].

\bibitem{Sadeghnezhad:2017hmr}
N.~Sadeghnezhad and K.~Nozari,
Phys.\ Lett.\ B {\bf 769} (2017) 134
doi:10.1016/j.physletb.2017.03.039
[arXiv:1703.06269 [gr-qc]].

\bibitem{Gorji:2017cai}
M.~A.~Gorji, S.~A.~Hosseini Mansoori and H.~Firouzjahi,
arXiv:1709.09988 [astro-ph.CO].

\bibitem{Zheng:2017qfs}
Y.~Zheng, L.~Shen, Y.~Mou and M.~Li,
JCAP {\bf 1708} (2017) no.08,  040
doi:10.1088/1475-7516/2017/08/040
[arXiv:1704.06834 [gr-qc]].

\bibitem{Matsumoto:2015wja}
J.~Matsumoto, S.~D.~Odintsov and S.~V.~Sushkov,
Phys.\ Rev.\ D {\bf 91} (2015) no.6, 064062
doi:10.1103/PhysRevD.91.064062
[arXiv:1501.02149 [gr-qc]].

\bibitem{Bouhmadi-Lopez:2017lbx}
M.~Bouhmadi-L\'opez, C.~Y.~Chen and P.~Chen,
arXiv:1709.09192 [gr-qc].

\bibitem{Myrzakulov:2015kda}
R.~Myrzakulov, L.~Sebastiani, S.~Vagnozzi and S.~Zerbini,
Class.\ Quant.\ Grav.\ {\bf 33} (2016) no.12, 125005
doi:10.1088/0264-9381/33/12/125005
[arXiv:1510.02284 [gr-qc]].


\bibitem{Oikonomou:2016fxb}
V.~K.~Oikonomou,
Int.\ J.\ Mod.\ Phys.\ D {\bf 25} (2016) no.07, 1650078
doi:10.1142/S0218271816500784
[arXiv:1605.00583 [gr-qc]].

\bibitem{Oikonomou:2016pkp}
V.~K.~Oikonomou,
Mod.\ Phys.\ Lett.\ A {\bf 31} (2016) no.33, 1650191
doi:10.1142/S0217732316501911
[arXiv:1609.03156 [gr-qc]].

\bibitem{Nojiri:2016ppu}
S.~Nojiri, S.~D.~Odintsov and V.~K.~Oikonomou,
Class.\ Quant.\ Grav.\ {\bf 33} (2016) no.12, 125017
doi:10.1088/0264-9381/33/12/125017
[arXiv:1601.07057 [gr-qc]].

\bibitem{Odintsov:2016imq}
S.~D.~Odintsov and V.~K.~Oikonomou,
Astrophys.\ Space Sci.\ {\bf 361} (2016) no.7, 236
doi:10.1007/s10509-016-2826-9
[arXiv:1602.05645 [gr-qc]].

\bibitem{Ade:2015lrj}
P.~A.~R.~Ade {\it et al.} [Planck Collaboration],
Astron.\ Astrophys.\  {\bf 594} (2016) A20
doi:10.1051/0004-6361/201525898 [arXiv:1502.02114 [astro-ph.CO]].

\bibitem{Array:2015xqh}
P.~A.~R.~Ade {\it et al.} [BICEP2 and Keck Array Collaborations],
Phys.\ Rev.\ Lett.\  {\bf 116} (2016) 031302
doi:10.1103/PhysRevLett.116.031302
[arXiv:1510.09217 [astro-ph.CO]].

\bibitem{Hirano:2017zox}
S.~Hirano, S.~Nishi and T.~Kobayashi,
JCAP {\bf 1707} (2017) no.07,  009
doi:10.1088/1475-7516/2017/07/009
[arXiv:1704.06031 [gr-qc]].

\bibitem{Nojiri:2016vhu}
S.~Nojiri, S.~D.~Odintsov and V.~K.~Oikonomou,
Phys.\ Rev.\ D {\bf 94} (2016) no.10,  104050
doi:10.1103/PhysRevD.94.104050
[arXiv:1608.07806 [gr-qc]].

\end{thebibliography}
\end{document}